\documentstyle[psfig,aps]{revtex}

\begin{document}

\title{Testing for General Dynamical Stationarity with a Symbolic Data Compression Technique}
\author{Matthew B. Kennel}
\address{
Institute for Nonlinear Science\\
University of California, San Diego\\
La Jolla, CA 92093-0402}

\author{Alistair I. Mees\thanks{Permanent address: Centre for Applied
    Dynamics and Optimization, The University of Western Australia,
    Nedlands, Perth 6907, Western Australia.}}
\address{
Isaac Newton Institute,\\
Cambridge University,\\
Cambridge CB3 0EH, England.}

\maketitle

\begin{abstract}
  
  We construct a statistic and null test for examining the {\em
    stationarity} of time-series of discrete symbols: whether two data
  streams appear to originate from the same underlying unknown
  dynamical system, and if any difference is statistically
  significant.  Using principles and computational techniques from the
  theory of data compression, the method intelligently accounts for
  the substantial serial correlation and nonlinearity found in
  realistic dynamical data, problems which bedevil naive methods.
  Symbolic methods are computationally efficient and robust to noise.
  We demonstrate the method on a number of realistic experimental
  datasets.
%
\end{abstract}

\section{Background}

Symbolic methods have been used in the study of dynamical systems from
the earliest days, most notably Kolmogorov and Sinai's~\cite{kolsin}
use of metric entropy as a dynamical invariant, which spawned a
significant mathematical industry in symbolic dynamics.
Fraser~\cite{fraser} applied information theoretical concepts to
construct useful algorithms and criteria for time-delay embeddings.

Stationarity, the notion that a system may be modeled well without
time as an explicit parameter, is a prerequisite for the vast majority
of nonlinear data analysis techniques.  Only recently has there been
some effort in constructing useful hypothesis tests suitable for
realistic chaotic and nonlinear dynamical data.\cite{statiotests} In
this work, we advocate a symbolic approach, on account of
computational ease, and the connection to well-studied and powerful techniques 
of data compression heretofore rarely used in the physics literature which
justify our statistical assumptions. 

Comparing information in the symbolic dynamics observed from time
series has a variety of uses besides stationarity
tests~\cite{Mees-Mason-Judd,Mees-Kennel}.  One that we mention here is a
'time-reversibility' test: is the symbol stream statistically the same
as its time-reversed version?  This is important because Gaussian
linear processes produce time-reversible time series, thus rejecting
time-reversibility in an observed symbol stream implies that the data
cannot be from that sort of process.

\section{The method} 
We have a series of symbols, either quantized from continuous-valued
observations or directly measured, observed in discrete time: $s_1,
s_2, s_3, \ldots s_N$, each symbol from some alphabet $A$ re-expressed
as integers $s \in \left\{1,2,\ldots,|A|\right\}$.  In symbolic data
analysis, the distribution of multi-symbol words provides information
about time-dependent structure and correlation, just as, with
continuous nonlinear data, time-delay embedding provides a vector
space revealing dynamical information.

A first attempt at a stationarity test would be to apply the classical
$\chi^2$-test to observed counts of distinct multi-symbol words
observed in (say) the front and back halves of the data.  Unfortunately,
the
assumption underlying this inference---that each datum is randomly and
{\em independently} drawn from some distribution ---is not true in
realistic dynamical data.  Short time correlations in physical data
strongly couple symbols near in time; thus naive application of
such tests fail miserably, usually towards spurious rejection.
Indeed, dynamical dependence makes it difficult to construct a
proper statistical null test for any hypothesis which allows chaotic
or general nonlinear data in the null class, and few examples of this
sort exist.

This work proposes a test procedure which quantifies whether two
observed symbol streams have ``the same dynamics'' and its statistical
significance, even in the presence of serial correlation and
dependence.  The algorithm is computationally rapid and does not
require Monte Carlo simulation.  There are two phases: construction of
a symbolic predictive model, and the evaluation of a combination of
classical statistics, this time on data constructed to be nearly
independent.

A model based on an universal data compression algorithm factors out
learnable dependence.  Good compressors learn the conditional
dependencies of symbols characteristic of the source process, thus
less new information need be transmitted to reproduce the input data,
assuming the decompresser can reconstruct the same model using the
transmitted symbols.  Fundamental results of information
theory~\cite{shannon,cover} require that optimally compressed data are
independent: this is the central theoretical justification for our
subsequent application of classical statistical inference, and we feel
one of the most useful concepts outside the specific application
presented here.

Our model for the symbolic dynamics is a ``context tree'': the recent
symbols in the stream themselves define the state, known here as the
{\em context}; contexts are analogous to the states reconstructed by
time-delay embedding in conventional nonlinear dynamical analysis.
Context tree modeling is a prominent contemporary development in the
data compression field.  We describe
elsewhere~\cite{Mees-Mason-Judd,Mees-Kennel} other applications to
nonlinear dynamics but in the present paper we specialize to
stationarity testing.

The tree structure accumulates the statistics of observed symbol
vectors down to maximum depth $d$, with distinct branches
corresponding to distinct symbols of alphabet $A$ which occurred at
prior times.  The top node corresponds to no history, the first $|A|$
nodes a one-dimensional reconstruction of the most recent symbol,
their $|A|^2$ descendents a 2-dimensional reconstruction of the two
most recent and so forth. Naturally one only constructs the non-empty
nodes.

Each node stores $|A|$ integers which record the distribution of every
symbol $s_{t+1}$ that occurred immediately after its particular
context $C$.  From this information we can estimate the conditional
probability for seeing the next symbol at every step:
$\hat{P}(s_{t+1}|C)$.  Observed data provides a data-based estimate of
the emission, and thus, transition, probabilities.



%
%
%
In principle we can build a context tree to arbitrary depth: the most
recently seen symbol can have a context that goes back to the start of
the data stream.  It is possible to construct such a tree efficiently
(in time and space linear in the number of
data~\cite{Willems:extensions})
but it is not useful to use the leaves directly to estimate
conditional probabilities, since each leaf will have only one
observation recorded.  We need a balance between greater depth, which
will reveal more structure, and larger numbers of observations at the
nodes, which will give greater robustness against noise: the
ubiquitous over-fitting issue.  There are various methods of either
selecting a specific subtree~\cite{rissanen} or
blending~\cite{Willems:weighted-context-trees} different subtrees, all
of which have varying
compression properties.  We use a state selection algorithm which 
is quite
convenient for our stationarity test.  We do not claim it is
necessarily the state-of-the-art compressor, though its performance
usually appears to be competitive on physical time-series data. 

At each time step $t$, the time series history so far selects a set of
possible contexts, or ``excited nodes'' of the tree, namely all nodes
reachable by following branches which match the symbol history.  We
use an additional criterion to select from among the the excited nodes
the ``encoding node'' at time $t$, which contains the estimated
$\hat{P}(s_{t+1})$ to be employed.  If one were literally compressing
data, one would feed the successive estimated $\hat{P}$ for the actual
symbols $s_{t+1}$ into an arithmetic coder, a well-studied algorithmic
device~\cite{cover} which generates the output bit stream with
total code length at most two bits greater than $-\sum_t \log_2
\hat{P}(s_{t+1})$.  At a node with observed counts $c_i$, we use the
Krichevsky-Trofimov~\cite{Shtarkov:multialphabet} estimator,
$$\hat{P}(j) =(c_j + \frac{1}{2})/\sum_{i} (c_i+ \frac{1}{2})$$
 for symbol value $1
\le j \le |A|$.  $\hat{P}$ must be estimated before the new future
symbol is included in the counts, because the decoder must be able to
reconstruct the symbol statistics from previously processed input at
every stage.  There is no separate model or dictionary sending step.
In our application, this is not essential and it is possible to
envisage batch encoders, which differ in detail but not in principle
from what we describe here.

Our method of encoding selects a specific encoding node for any symbol
using a ``predictive minimum description length (MDL)'' principle due
to Rissanen and later improved by others~\cite{rissanen}.  Each node
stores a differential code length $\Delta = L_p - L_s$, the difference
in code lengths which would have been emitted had the symbol been
encoded using estimated distributions $\hat{P}$ at itself ($L_s$), or
code length $L_p$ estimated using the parent's counts.  One descends
the excited nodes from the top down, computing $\sum \Delta$, summing
over {\em all} of the current node's children's value of $\Delta$.
When this sum first becomes negative (assigning $\Delta=0$ to
nonexistent child nodes) we have found the encoding node.  The
selection process has found the shortest matching context for which it
would have been cheaper to have encoded past data using that context
than longer matching contexts.


%
%

After encoding the current symbol, one updates $\Delta$ for
each excited node:
\begin{equation}
 \Delta \leftarrow \Delta
+(-\log_2 \hat{P}_p(s_{t+1}) + \log_2 \hat{P}_s(s_{t+1})),
\end{equation} the
first expression using the node's parent's counts and the second using
the node's own counts.  Note that $\Delta$ is a quantity maintained
for each node independent of any previous choice of encoding node.
Lastly, the conditional counts $c_j$ for all excited nodes are
appropriately incremented with the new symbol, and new branches of the
tree grafted for histories never seen before.  It is important that
the three phases be carried out in this particular sequence, repeating
all three for each new symbol to encode.

Notice that, unlike fixed Markov modeling or the simplest version of
time-delay
embedding, the number of past symbols which contribute to predicting
the future in a context tree is not uniform.  Some past histories need
to be deeply
examined because there, long-past history influences the future
significantly, whereas for other past histories, there is less need,
either because future evolution is more unpredictable or there has
been less information previously observed regarding those histories.
In this respect, contexts are like ``variable
embeddings''~\cite{judd-mees}.

\section{Stationarity test}

After encoding all the symbols we carry out the stationarity test.
The overall goal, answering the question ``do two data sets appear to
arise from the same
underlying dynamical system'', translates to combining hypothesis tests
performed at each encoding node regarding whether the encodings
observed for both data sets (the distribution of future symbols
actually encoded) could have come from a single underlying probability
distribution $P(k)$, and if any apparent difference is statistically
significant.  At encoding contexts, we may use standard tests because
these events ought to be nearly independent using a good compression
algorithm.  (A perfectly compressed data stream would be
indistinguishable from a stream of independent
random Bernoulli bits under any statistical test.) 

At any node $n$, we have recorded the frequency with which symbol $k$
was encoded here, $e_{k;1}$ in the first set and $e_{k;2}$ in the
second.  (Note that $e_k \ne c_k$, the latter accumulating frequencies
whenever a context was excited.) Assuming independence, the statistic
\begin{equation}
\chi^2 = \sum_{k=1}^{|A|} \frac{\left( R^{1/2} e_{k;1} - 
        R^{-1/2} e_{k;2} \right)^2 }{e_{k;1}+e_{k;2}}
\end{equation}
with $R = \sum e_{k;2} / \sum e_{k;1}$ follows the standard $\chi^2$
distribution with $|A| -1$ degrees of freedom under the null
hypothesis that both empirical probability distributions came from the
same underlying distribution\cite{chicaveat}. Given the value
of $\chi^2$ and the degrees of freedom, standard numerical algorithms
provide a likelihood $L$ asymptotically uniform $L \in (0,1)$ under
the null.  Small values of $L$ reject the null at the given
significance level, e.g. $\L < 0.01$.

The asymptotic distribution of the $\chi^2$ test used in the
computation of $L$ becomes increasingly inaccurate as the number of
observations decreases.  Thus for $\sum e_k < 75$ (an arbitrary cutoff)
we instead use a combinatorial test for differences in proportions,
called {\em Fisher's exact test}.  As the test is much simpler in the
$2\times 2$ case, we keep the observation for the most frequent symbol
(bin $m$ which achieves $\max (e_{m;1}+e_{m;2})$) and merge the others
into $e_{o;1}, e_{o;2}$, resulting in four quantities conventionally
expressed in a ``contingency table'', with cumulative row and column
sums:
$$
\begin{array}{c|c|c}
 e_{m;1} & e_{o;1} & n_1  \\
e_{m;2} & e_{o;2} & n_2  \\
\hline
 n_m & n_o & N 
\end{array}
$$
Under the null that the difference in proportions between $m$ and
$o$ counts is independent of being in set 1 and 2, the probability for
seeing any particular table with the given marginal sums is:
$$p_T = n_m! n_o! n_1! n_2! / (e_{m;1}! e_{m;2}! e_{o;1}! e_{o;2}!
N!).$$
One directly enumerates all tables with the given observed
marginals (only a 1-d sum for a $2 \times 2$ table) and sums $p_T$ for
every table with a difference in proportions at least as great as that
observed\cite{fishercaveat}, resulting in a likelihood $L$ for
accepting the null hypothesis at this node.

We combine these $M$ likelihoods, each measuring some aspect of of the
same null hypothesis, into a single overall test. Under the null, the
quantity
\begin{equation}
  X^2 = \sum_{k=1}^{M} -2 \ln L_k
\end{equation}
is $\chi^2$ distributed with $2M$ degrees of freedom, from which we
compute our final ${\cal L}$, again uniform in $(0,1)$ under the null.
Especially small values of ${\cal L}$ imply a small likelihood that
this level of difference would have been observed had the two symbol
datasets been generated by the same underlying dynamical process.
This completes our desired test procedure.

\section{Applications}

\begin{figure}
\centerline{\psfig{file=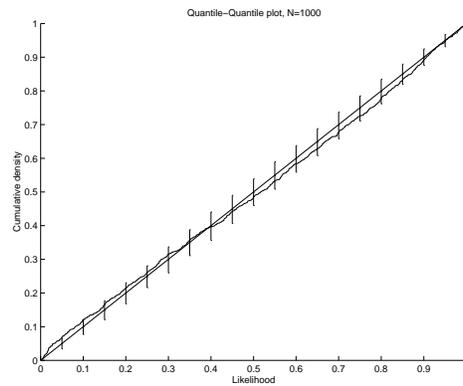,height=2in}}
\caption{Quantile-quantile plot of ${\cal L}$ under the null hypothesis.
  The observed values of ${\cal L}$ are sorted and plotted vs their
  normalized index $(i+1)/1001$.  Asymptotically the curve should
  approach the diagonal under the null.  Bars are $\pm$ two standard deviations
for 100 samples of 1000 uniform deviates $\in [0,1]$ processed similarly.}
\label{fig:lorstatio}
\end{figure}

We first test the accuracy of the null hypothesis. We produced an
ensemble of 1000 time series from the $x$ coordinate of the ``Lorenz
84'' attractor: a tiny geophysical model with attractor dimension $d
\approx 2.5$~\cite{lorenz84}.  Figure \ref{fig:lorstatio} shows the
distribution of ${\cal L}$ comparing the first and second halves of
each set, demonstrating ${\cal L}$ is close to uniform $\in (0,1)$.
This is a stringent requirement and shows the success of our
independence assumption, as it is difficult to get a high-quality null
distribution with complicated arbitrarily correlated chaotic data in
the null class.  With this number of data, the test is also quite
powerful.

\begin{figure}
\centerline{\psfig{file=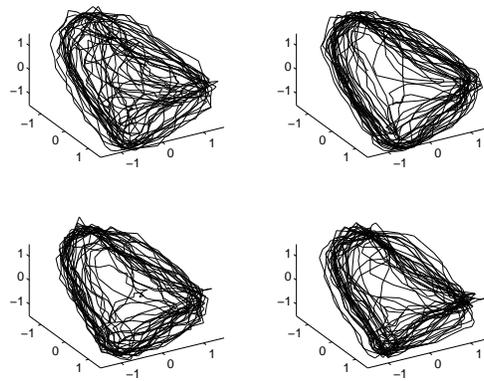,height=2in}}
\caption{Phase space plots of the differential pressure signal from a fluidized
bed reactor.  Three are from the same parameters, one is different.}
\label{fourbeds}
\end{figure}

\begin{figure}
  \centerline{\psfig{file=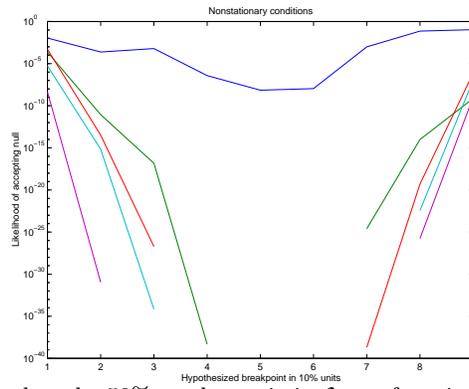,height=2in}}
\caption{One
  fluidized bed system altered at the 50\% mark, statistic ${\cal L}$
  as a function of hypothesized breakpoint and symbolic alphabet
  precision.  Curves for $|A|>2$ numerically underflowed to ${\cal L}
  = 0$ toward the center and aren't plotted.  Null hypothesis
  emphatically rejected.}
\label{bednulltest}
\end{figure}

We demonstrate discrimination power with a set of pressure data from
an experimental model of a ``fluidized bed reactor'' \cite{bedprl}.
This experimental system consists of a vertical cylindrical tube of
granular particles excited from below by an externally input gaseous
flow.  In some parameter regimes (``slugging''), the particles exhibit
complex motion which appears to be a combination of collective
low-dimensional bulk dynamics and small-scale high-dimensional
turbulence of the individual particles~\cite{bedprl}. The observed
variable was an azimuthally averaged pressure difference between two
vertically separated taps.  Figure~\ref{fourbeds} shows portions of
time-delay embedding of orbits sections of the dataset taken at the
same experimental parameters, and one when the flow was boosted by
5\%.  The change in the attractor is rather subtle and difficult to
reliably diagnose by eye.  Figure~\ref{bednulltest} shows the result
of calculating ${\cal L}$ on a data set whose flow was increased at
the midpoint.  As the alphabet size increased and hypothesized
breakpoint approached the true value of 50\%, the strength of the
rejection increased, ${\cal L} \rightarrow 0$. Even the binary
alphabet case showed a significant rejection of the null.  On data
taken in stationary conditions ${\cal L}$ fluctuates randomly in
$(0,1)$, as expected.

\begin{figure}
\centerline{\psfig{file=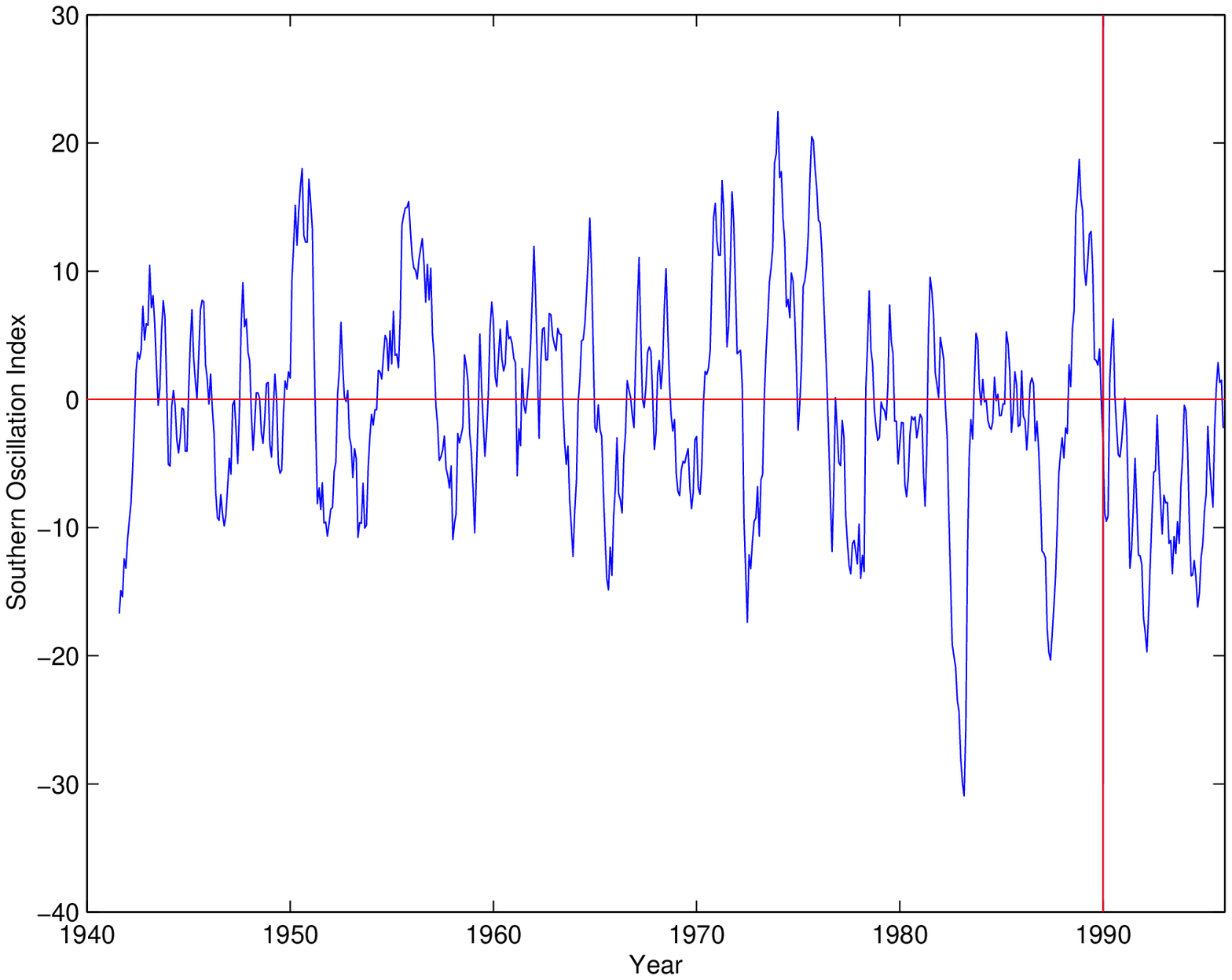,height=2in}}
\caption{Three month moving average of the Southern Oscillation Index,
  the normalized pressure difference between Tahiti and Darwin, Australia.
  Strongly negative values correspond to El Nino events.  Is the extended
  negative period from mid 1990 through 1995 especially anomalous?}
\label{fig:soi}
\end{figure}

The Southern Oscillation Index, the normalized pressure difference
between Tahiti and Darwin, is a proxy for the El Nino Southern
Oscillation, as ocean temperature influences atmospheric dynamics. The
period from mid 1990 to 1995 exhibited an anomalously sustained period
of El Nino-like conditions (Fig.~\ref{fig:soi}), perhaps indicative
of global climate change.  One statistical analysis~\cite{elnino1}
found such an anomaly quite unlikely assuming stationarity, but
another group~\cite{elnino2} used a different analysis and found it
significantly more likely to be a chance fluctuation.  Both papers
used traditional linear forecasting models, with the difference
centered around an auto-correlation based correction for serial
correlation to arbitrarily reduce the degrees of freedom.  We applied
our algorithm to the 3-month moving average SOI (binary symbolized)
testing the 5.4 year period in question against the rest of the
series, with a resulting ${\cal L} \approx 0.01$, meaning that one
would expect to see a region this anomalous by chance every 540 years.
The result is closer to those of~\cite{elnino2} than~\cite{elnino1}
but we certainly do not want to take any particular position regarding
climate; rather, we wish to point out an application for our method where
correcting for serial correlation automatically is useful.

Recent work has successfully used a distance in the symbolic space to
fit unknown parameters of a physically motivated continuous model to
observed data, including substantial observational and dynamic noise
all in one framework, the situation where fitting models directly is
difficult or unreliable.  Tang et al~\cite{tangbrowntracy} first
proposed minimizing over free parameters the difference between an
observed distribution of symbol words and that produced by
discretizing some proposed model's output.  Daw et
al~\cite{dawfinkenn} successfully used employed this technique to fit
experimental internal combustion engine measurements to a
low-dimensional dynamical model.  The optimization target was a
Euclidean norm in \cite{tangbrowntracy} and a chi-squared distance in
\cite{dawfinkenn}.  On account of serial correlation, a true
hypothesis test confirming or rejecting the apparent compatibility of
observed data to best-fitting model was not possible in those works.
We feel that our current method ought to provide a more intelligent
and less ad hoc optimization goal, either by maximizing average ${\cal
  L}$ or perhaps minimizing the code length of the physical model's
output, encoded using the symbolic model learned from the observed
data.

\newpage
\begin{figure}
\centerline{\psfig{file=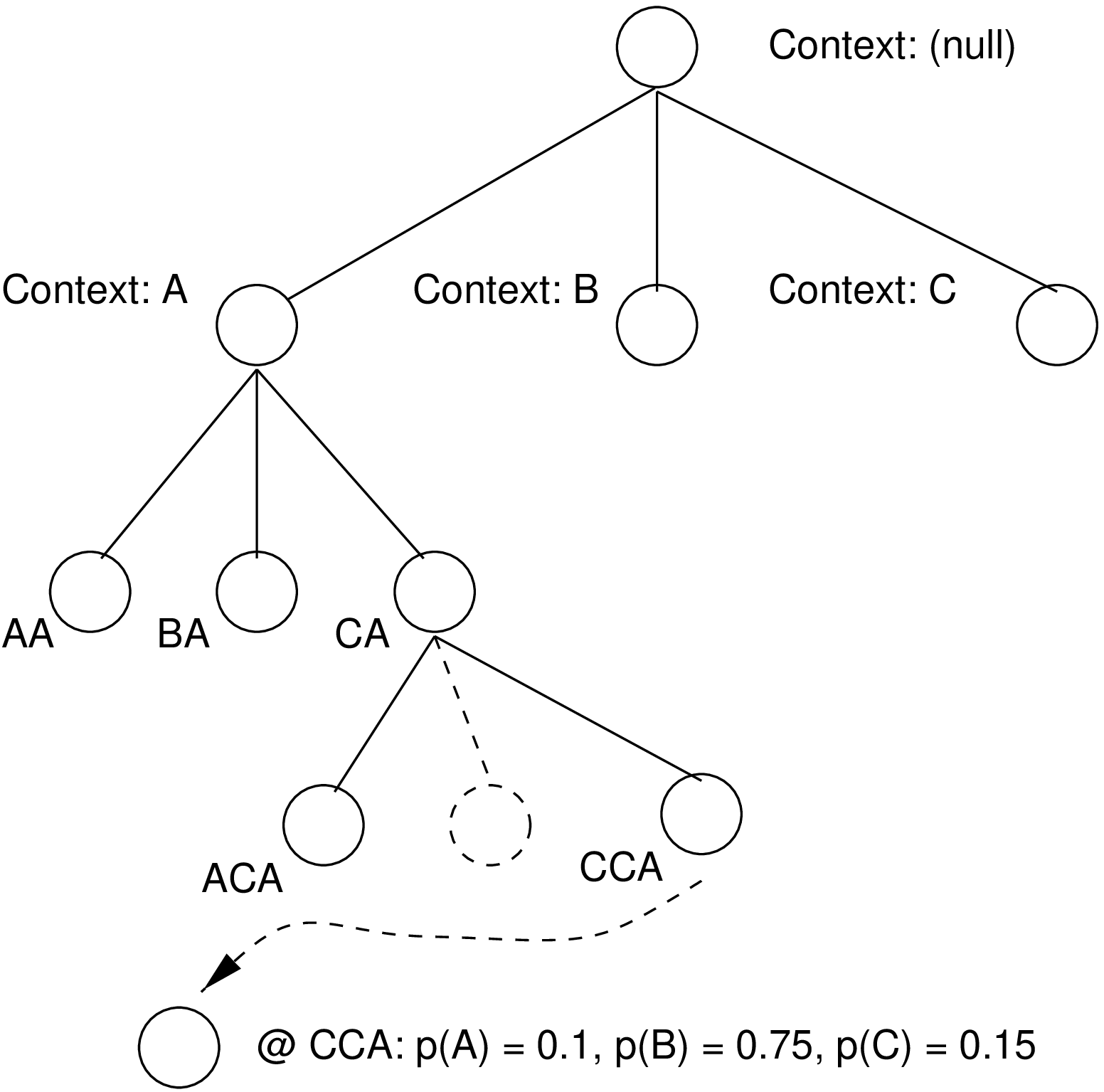,height=2in}}
\caption{Example of a small context tree for a 3 symbol alphabet.  The
  dotted node indicates that not all children necessarily exist, depending
  on the past symbols observed.  Every node stores an estimated
  probability distribution, shown here for one node.}
\end{figure}

\begin{references}
\bibitem{chicaveat} The $\chi^2$ analytics degrade for small bin
  counts. For those bins, we merge any symbols whose expected
  count--for either set one or two--is less than five, reducing the
  degrees of freedom appropriately.  If, after merging, there remain
  fewer than two symbols passing this criterion, then this node is
  wholly excluded.
\bibitem{fishercaveat} In this discrete case, the sum will encounter
  tables exactly as likely as the observed one (such as the observed
  table itself); the summed $p_T$ for these tables is weighted by
  a uniform random deviate $r \in [0,1)$.
\end{references}
\end{document}